# Generation of Higher Dimensional Modal Entanglement Using a Three Waveguide Directional Coupler

Divya Bharadwaj[1], K Thyagarajan[1], Michal Karpinski[2] and Konrad Banaszek[3]
[1]*Department of Physics, IIT Delhi, New Delhi 110016, India*
[2]*Clarendon Laboratory, University of Oxford, Parks Road, Oxford, OX1 3PU, UK*
[3]*Faculty of Physics, University of Warsaw, Warsaw, Poland*

In this paper, we propose a method for the generation of higher dimensional modal entanglement through type II spontaneous parametric down conversion process using a three waveguide directional coupler in a periodically poled lithium niobate substrate. We show that by a proper design, it is possible to achieve an output state of two photons occupying three different spatial modes. The advantage of using such waveguide structure is its flexibility and the design space availability to achieve desired characteristics of the photon pairs generated in the down conversion process.

## I. INTRODUCTION

Entangled photons have become one of the most important ingredients in the field of quantum computing, quantum cryptography [1, 2] and quantum teleportation [3]. Generation and manipulation of entangled states in different degrees of freedom such as polarization, spatial and spectral have been extensively studied in the literature for various applications in quantum communication protocols such as quantum key distribution (QKD) [4-5], entanglement swapping [6], quantum super dense coding [7] etc. Most of these studies involve two dimensional entanglements. Recently, focus has shifted towards studying higher dimensional entanglement i.e. entanglement in more than two modes. Quantum states with higher dimensional entanglement provide larger information capacity and increased noise threshold in comparison to entangled system in two dimensions [8, 9]. In the literature several ways have been proposed to generate higher dimensional entanglement such as higher dimensional entanglement of orbital angular momentum (OAM) of photons [10, 11], higher dimensional time- bin entanglement [12] etc.

In this paper, we propose a method to generate a higher dimensional mode entangled photon pairs through type II spontaneous parametric down conversion (SPDC) process in a three waveguide directional coupler. It is shown that by an appropriate design it is possible to achieve an output state of two photons occupying three different spatial modes. This structure has an advantage of flexibility and the design space availability to achieve desired characteristics of the photon pairs generated in the down conversion process in comparison to the multimode single waveguide structure. In addition to that, we also show that with the same structure and quasi phase matched (QPM) grating, it is possible to get an output state of two photons occupying four different possible combinations of three spatial modes. This state corresponds to partial high dimensional entanglement. The partially entangled states find applications in probabilistic teleportation [13, 14].

## II. PRINCIPLE

We consider a three waveguide directional coupler consisting of three identical single mode waveguides each of width '$a$' separated by a distance '$d$' such that the three waveguide coupler supports three normal guided modes (two symmetric and one antisymmetric) referred to as 0, 1 and 2 as shown in Figure 1. We represent the effective indices of the modes at different frequencies by $n_{\alpha m}$ where $\alpha = p, s, i$ for pump, signal and idler respectively and $m = 0, 1, 2$ for the three modes.

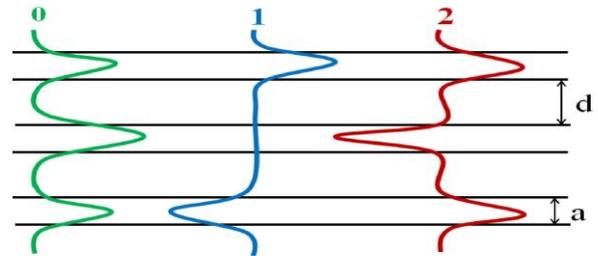

FIG. 1: (Color online) Schematic diagram of an array of three identical single mode waveguides and the field distributions of the three normal modes.

We consider generation of photon pairs using degenerate parametric down conversion with type II phase matching with a horizontally (H) polarized pump and consider the signal to be horizontally (H) polarized and the idler to be vertically (V) polarized. Using an adiabatic evolution of the waveguide structure from a single waveguide to the three coupled waveguides (see Figure 1), at the pump wavelength it is possible to excite the H-polarization of either the fundamental symmetric mode or the first excited anti symmetric mode of the three waveguide region.

The parametric down conversion process from the pump photon to signal idler pair would depend on the phase matching or quasi phase matching condition that is satisfied as well as on the overlap integral between the interacting pump, signal and idler modes. We consider two cases:

### A. Pump photon in the fundamental symmetric mode

We first assume the pump to be in the fundamental symmetric mode of the three waveguide coupler. Due to parity conservation, the H-polarized fundamental symmetric pump mode can only down convert to any of the following combination of pairs of modes of signal and idler of orthogonal polarization: $H_{s0}V_{i0}$; $H_{s1}V_{i1}$; $H_{s0}V_{i2}$; $H_{s2}V_{i0}$ and $H_{s2}V_{i2}$, where H and V correspond to horizontal and vertical polarization states, subscripts $s$ and $i$ correspond to signal and idler and the integer corresponds to the mode number. The remaining combinations of symmetric-anti symmetric pairs of signal idler pairs are not allowed due to parity conservation requirement [15-17].

In order to achieve higher dimensional entangled photon pairs, we need to properly design the waveguides and their separation so that three of the above mentioned processes occur with the same probability. In order to show that this is possible, we choose the following three possibilities of down conversion process: $H_{s0}V_{i0}$; $H_{s1}V_{i1}$ and $H_{s2}V_{i2}$; these have been chosen since in the case of the three waveguide directional coupler all the three processes are characterized by almost the same efficiency. In such a case the output is expected to be an entangled state given by:

$$|\psi_0\rangle = i\int d\omega [C_{01}|H_{s0},V_{i0}\rangle + C_{02}|H_{s1},V_{i1}\rangle + C_{03}|H_{s2},V_{i2}\rangle] \quad (1)$$

where, $C_{01}$, $C_{02}$ and $C_{03}$ are constants which are determined by the field overlap integral and the phase matching function. These three processes require quasi phase matching (QPM) nonlinear gratings with spatial frequencies given by:

$$K_{01} = \frac{2\pi}{\Lambda_{01}} = 2\pi\left(\frac{n_{p0}}{\lambda_p} - \frac{n_{s0}}{\lambda_s} - \frac{n_{i0}}{\lambda_i}\right) \quad (2a)$$

$$K_{02} = \frac{2\pi}{\Lambda_{02}} = 2\pi\left(\frac{n_{p0}}{\lambda_p} - \frac{n_{s1}}{\lambda_s} - \frac{n_{i1}}{\lambda_i}\right) \quad (2b)$$

$$K_{03} = \frac{2\pi}{\Lambda_{03}} = 2\pi\left(\frac{n_{p0}}{\lambda_p} - \frac{n_{s2}}{\lambda_s} - \frac{n_{i2}}{\lambda_i}\right) \quad (2c)$$

where, $\lambda_{p(s,i)}$ is the pump (signal, idler) wavelength, $\Lambda_{0j}$ represents grating period for the $j^{th}$ process.

Thus when the pump photon (assumed to be horizontally polarized and in 0 spatial mode) is incident on the waveguide, it will down convert via any one of the following three (j=1, 2, 3) processes:

$$H_{p0} \to H_{s0} + V_{i0}$$
$$H_{p0} \to H_{s1} + V_{i1}$$
$$H_{p0} \to H_{s2} + V_{i2}$$

The output state will be maximally entangled if the values of coefficients $C_{01}$, $C_{02}$ and $C_{03}$ (in Eq. (1)) become equal. Since these coefficients depend upon the overlap integrals, effective indices of the interacting modes at the pump, signal and idler wavelengths and also the phase matching function, it will be shown in Sec. IV that by an appropriate design of the waveguide and their separation it is possible to achieve this condition and thus obtain a higher dimensional entangled state. We will also show in Sec. IV that by an appropriate choice of the waveguide parameters all the spatial frequencies required as per Eq. (2) can be made very close to each other i.e. $K_{01} \approx K_{02} \approx K_{03} \approx K$, thus providing the possibility of all the processes using a single QPM grating.

### B. Pump photon in the first excited anti symmetric mode

We next consider the pump to be in the first excited mode of the three waveguide coupler. In this case the *H*-polarized first excited anti symmetric pump mode can only down convert to any of the following combination of pairs of modes of signal and idler of orthogonal polarization: $H_{s0}V_{i1}$; $H_{s1}V_{i0}$; $H_{s1}V_{i2}$ and $H_{s2}V_{i1}$ due to parity conservation. In such a case the output is expected to be an entangled state given by:

$$|\psi_1\rangle = i\int d\omega [C_{11}|H_{s0},V_{i1}\rangle + C_{12}|H_{s1},V_{i0}\rangle + C_{13}|H_{s1},V_{i2}\rangle + C_{14}|H_{s2},V_{i1}\rangle] \quad (3)$$

These four processes require quasi phase matching (QPM) nonlinear gratings with spatial frequencies given by:

$$K_{11} = \frac{2\pi}{\Lambda_{11}} = 2\pi\left(\frac{n_{p1}}{\lambda_p} - \frac{n_{s0}}{\lambda_s} - \frac{n_{i1}}{\lambda_i}\right) \quad (4a)$$

$$K_{12} = \frac{2\pi}{\Lambda_{12}} = 2\pi\left(\frac{n_{p1}}{\lambda_p} - \frac{n_{s1}}{\lambda_s} - \frac{n_{i0}}{\lambda_i}\right) \quad (4b)$$

$$K_{13} = \frac{2\pi}{\Lambda_{13}} = 2\pi\left(\frac{n_{p1}}{\lambda_p} - \frac{n_{s1}}{\lambda_s} - \frac{n_{i2}}{\lambda_i}\right) \quad (4c)$$

$$K_{14} = \frac{2\pi}{\Lambda_{14}} = 2\pi\left(\frac{n_{p1}}{\lambda_p} - \frac{n_{s2}}{\lambda_s} - \frac{n_{i1}}{\lambda_i}\right) \quad (4d)$$

Thus when the pump photon (assumed to be horizontally polarized and in spatial mode 1) is incident on the waveguide, it will down convert via any one of the following four ($j=1, 2, 3, 4$) processes:

$$H_{p1} \to H_{s0} + V_{i1}$$
$$H_{p1} \to H_{s1} + V_{i0}$$
$$H_{p1} \to H_{s1} + V_{i2}$$
$$H_{p1} \to H_{s2} + V_{i1}$$

The output state will be maximally entangled if the values of coefficients $C_{11}$, $C_{12}$, $C_{13}$ and $C_{14}$ (in Eq. (3)) become equal. Since these coefficients depend upon the overlap integrals, effective indices of the interacting modes at the pump, signal and idler wavelengths and also the phase matching function, it will be shown in Sec. IV that by an appropriate design of the waveguide and their separation it is possible to achieve this condition and thus obtain a partial high dimensional entangled state. We will also show in Sec. IV that by an appropriate choice of the waveguide parameters all spatial frequencies required can be made very close to each other i.e. $K_{11} \approx K_{12} \approx K_{13} \approx K_{14} \approx K$, thus providing the possibility of all the processes using a single QPM grating.

## III. ANALYSIS

We consider the three waveguide directional coupler in $z$-cut periodically poled lithium niobate (Figure 2) with the optic axis along $z$-axis so that the H-polarization (along the $y$-direction) corresponds to ordinary polarization while the V-polarization (along the $z$-direction) corresponds to the extraordinary polarization. Here the waveguide width and depth are assumed to be such that the three waveguide coupler supports only the 00, 10 and 20 modes. In order to shorten the symbols, we represent the 00,10 and 20 modes of channel waveguide by 0, 1 and 2 respectively.

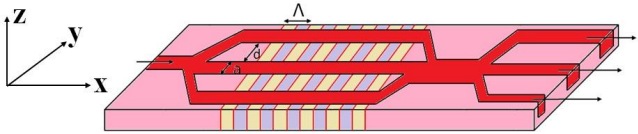

FIG. 2: (Color online) Schematic of the channel waveguide of three waveguide coupler having QPM grating of period $\Lambda$

In order to obtain the output down converted quantum state, we need to analyze the normal modes of the three waveguide directional coupler. As described in Appendix A, a standard approximate analysis can be used to obtain the propagation constants as well as the modal field distributions of the H- and V-polarized normal modes of the coupler.

The transverse field distributions $u_{p(s,i)}^{(m)}(\vec{r})$ of the three normal modes can be described by the following equation:

$$u_{p(s,i)}^{(m)}(\vec{r}) = Y_{p(s,i)}^{(m)}(y) Z_{p(s,i)}^{0}(z) \quad (5)$$

where, $m$ represents the mode number and the subscripts represent pump, signal and idler fields. The overall field is written as a product of the y-dependence and z-dependence; the waveguide geometry supports only the 0 order mode along the vertical direction. $Y_{p(s,i)}^{(m)}(y)$ is the electric field profile of pump (signal, idler) in the $m^{th}$ mode along $y$-direction, $Z_{p(s,i)}^{0}(z)$ is the electric field profile of pump (signal, idler) of fundamental mode along $z$-direction.

Now, the electric fields at pump, signal and idler corresponding to different modes are represented by the following equations [18]:

$$\vec{E}_p^{(m)} = u_p^{(m)}(\vec{r}) A_m \cos(\beta_p^{(m)} x - \omega_p t)$$
$$= \frac{1}{2} u_p^{(m)}(\vec{r}) A_m \left( e^{i(\beta_p^{(m)} x - \omega_p t)} + e^{-i(\beta_p^{(m)} x - \omega_p t)} \right) \quad (6)$$

$$\hat{E}_s^{(m)} = i \int d\omega u_s^{(m)}(\vec{r}) \sqrt{\frac{hc}{2\varepsilon_0 \lambda_s n_{sm}^2 L}} (\hat{a}_{sm} e^{i\beta_s^{(m)} x} - \hat{a}_{sm}^\dagger e^{-i\beta_s^{(m)} x}) \quad (7)$$

$$\hat{E}_i^{(m)} = i \int d\omega u_i^{(m)}(\vec{r}) \sqrt{\frac{hc}{2\varepsilon_0 \lambda_i n_{im}^2 L}} (\hat{a}_{im} e^{i\beta_i^{(m)} x} - \hat{a}_{im}^\dagger e^{-i\beta_i^{(m)} x}) \quad (8)$$

where, $L$ represents the length of interaction, $\beta_{p(s,i)}^{(m)} = 2\pi \frac{n_{p(s,i)m}}{\lambda_{p(s,i)}}$ is the propagation constant of the pump(signal, idler) in the $m^{th}$ mode, $h$ is the Planck's constant, $c$ is the speed of light and $\varepsilon_0$ is the free space permittivity.

We assume the pump to be described by a classical field as it is assumed to be strong. Here, $A_m$ is the amplitude of pump in the $m^{th}$ mode. In Eq. (7) and Eq. (8) $\hat{a}$ and $\hat{a}^\dagger$ represent the annihilation and creation operators of the generated signal $(s)$ and idler $(i)$ photons corresponding to the $m^{th}$ mode.

The second order nonlinear polarization in the medium is given by:

$$P_k^{NL} = 2\varepsilon_0 \sum_{l,m} d_{klm} E_l E_m \quad (9)$$

where, $E_l$ represents the $l^{th}$ component of the total electric field within the medium. The interaction Hamiltonian is given by:

$$\hat{H}_{int} = \int U dV \quad (10)$$

where, $U = -\int P_k dE_k$

In accordance with the interaction picture, the overall output state is given as:

$$|\psi\rangle = e^{-iH_{int}t/h} |0_s, 0_i\rangle \quad (11)$$

where, $|0_s, 0_i\rangle$ correspond to vacuum state i.e. no signal and idler photon.

### A. Pump is fundamental symmetric mode:

In this case, the electric field components are given by:

$$E_x = 0,$$
$$E_y = E_p^{(0)} + E_s^{(0)} + E_s^{(1)} + E_s^{(2)} \quad (12)$$
$$E_z = E_i^{(0)} + E_i^{(1)} + E_i^{(2)}$$

Using Eq. (9) and (12) in Eq. (10) and considering only the quasi phase matched processes since other terms are negligibly small as compared to the phase matched terms, we get interaction Hamiltonian as:

$$H_{int} = -4\varepsilon_0 \int d_{24} [E_p^{(0)} E_s^{(0)} E_i^{(0)} + E_p^{(0)} E_s^{(1)} E_i^{(1)}$$
$$+ E_p^{(0)} E_s^{(2)} E_i^{(2)}] dV \quad (13)$$

$$H_{int} = \int d\omega [C'_{01} \hat{a}_{s0}^\dagger \hat{a}_{i0}^\dagger e^{-i\omega_p t} + C'_{02} \hat{a}_{s1}^\dagger \hat{a}_{i1}^\dagger e^{-i\omega_p t}$$
$$+ C'_{03} \hat{a}_{s2}^\dagger \hat{a}_{i2}^\dagger e^{-i\omega_p t} + Hermitian\, Conjugate(H.C.)] \quad (14)$$

Using the interaction picture (Eq. (11)) we can obtain the overall output quantum state (neglecting the vacuum state contribution) as:

$$|\psi_0\rangle = i \int d\omega [C_{01} |H_{s0}, V_{i0}\rangle + C_{02} |H_{s1}, V_{i1}\rangle$$
$$+ C_{03} |H_{s2}, V_{i2}\rangle] \quad (15)$$

where,

$$C_{0j} = -t \frac{C'_{0j}}{\hbar}$$

$$= -\frac{(4cd_{24} A_0 t)}{\sqrt{\lambda_s \lambda_i}} \frac{I_{0j} \text{sinc}\left(\frac{\Delta k_{0j} L}{2}\right)}{N_{0j}} e^{-i\frac{\Delta k_{0j} L}{2}} \quad ; j=1, 2, 3$$

$$= F_0 \frac{I_{0j} \text{sinc}\left(\frac{\Delta k_{0j} L}{2}\right)}{N_{0j}} e^{-i\frac{\Delta k_{0j} L}{2}} = F_0 F_{0j}$$

where, $N_{01} = n_{s0} n_{i0}$, $N_{02} = n_{s1} n_{i1}$ and $N_{03} = n_{s2} n_{i2}$

$$F_0 = \frac{4cd_{24} A_0 t}{\sqrt{\lambda_s \lambda_i}} \text{ and } F_{0j} = \frac{I_{0j} \text{sinc}\left(\frac{\Delta k_{0j} L}{2}\right)}{N_{0j}} e^{-i\frac{\Delta k_{0j} L}{2}}$$

$\Delta k_{0j}$ is the phase mismatch given by:

$$\Delta k_{01} = \beta_p^{(0)} - \beta_s^{(0)} - \beta_i^{(0)} - K$$
$$\Delta k_{02} = \beta_p^{(0)} - \beta_s^{(1)} - \beta_i^{(1)} - K \quad (16)$$
$$\Delta k_{03} = \beta_p^{(0)} - \beta_s^{(2)} - \beta_i^{(2)} - K$$

$I_{0j}$ represent the overlap integrals between the pump, signal and idler of $j=1, 2,$ and $3$ processes, and are given by:

$$I_{01} = \int u_p^{(0)}(\vec{r}) u_s^{(0)}(\vec{r}) u_i^{(0)}(\vec{r}) dydz$$
$$I_{02} = \int u_p^{(0)}(\vec{r}) u_s^{(1)}(\vec{r}) u_i^{(1)}(\vec{r}) dydz \quad (17)$$
$$I_{03} = \int u_p^{(0)}(\vec{r}) u_s^{(2)}(\vec{r}) u_i^{(2)}(\vec{r}) dydz$$

The efficiency of the three down conversion processes for $j=1, 2,$ and $3$ are proportional to $|F_{0j}|^2$

### B. Pump is first excited antisymmetric mode:

In this case, the electric field components are given by:

$$E_x = 0,$$
$$E_y = E_p^{(1)} + E_s^{(0)} + E_s^{(1)} + E_s^{(2)} \quad (18)$$
$$E_z = E_i^{(0)} + E_i^{(1)} + E_i^{(2)}$$

Using Eq. (9), (10), (18) and considering only the phase matched processes as other terms are negligibly small as compared to the phase matched terms, we get interaction Hamiltonian as:

$$H_{int} = -4\varepsilon_0 \int d_{24} [E_p^{(1)} E_s^{(0)} E_i^{(1)} + E_p^{(1)} E_s^{(1)} E_i^{(0)}$$
$$+ E_p^{(1)} E_s^{(1)} E_i^{(2)} + E_p^{(1)} E_s^{(2)} E_i^{(1)}] dV \quad (19)$$

$$H_{int} = \int d\omega [C'_{11} \hat{a}_{s0}^\dagger \hat{a}_{i1}^\dagger e^{-i\omega_p t} + C'_{12} \hat{a}_{s1}^\dagger \hat{a}_{i0}^\dagger e^{-i\omega_p t} +$$
$$C'_{13} \hat{a}_{s1}^\dagger \hat{a}_{i2}^\dagger e^{-i\omega_p t} + C'_{14} \hat{a}_{s2}^\dagger \hat{a}_{i1}^\dagger e^{-i\omega_p t} + H.C.] \quad (20)$$

Using the interaction picture (Eq. (11)) the output state is given as:

$$|\psi_1\rangle = i \int d\omega [C_{11} |H_{s0}, V_{i1}\rangle + C_{12} |H_{s1}, V_{i0}\rangle$$
$$+ C_{13} |H_{s1}, V_{i2}\rangle + C_{14} |H_{s2}, V_{i1}\rangle] \quad (21)$$

where,

$$C_{1j} = -t\frac{C'_{1j}}{\hbar}$$

$$= -\frac{(4cd_{24}A_1 t)}{\sqrt{\lambda_s \lambda_i}} \frac{I_{1j}\mathrm{sinc}\left(\frac{\Delta k_{1j}L}{2}\right)}{N_{1j}} e^{-i\frac{\Delta k_{1j}L}{2}} \quad ; j=1, 2, 3, 4$$

$$= F_1 \frac{I_{1j}\mathrm{sinc}\left(\frac{\Delta k_{1j}L}{2}\right)}{N_{1j}} e^{-i\frac{\Delta k_{1j}L}{2}} = F_1 F_{1j}$$

$N_{11} = n_{s0}n_{i1}$, $N_{12} = n_{s1}n_{i0}$, $N_{13} = n_{s1}n_{i2}$ and $N_{14} = n_{s2}n_{i1}$

$$F_1 = \frac{4cd_{24}A_1 t}{\sqrt{\lambda_s \lambda_i}} \text{ and } F_{1j} = \frac{I_{1j}\mathrm{sinc}\left(\frac{\Delta k_{1j}L}{2}\right)}{N_{1j}} e^{-i\frac{\Delta k_{1j}L}{2}}$$

$\Delta k_{1j}$ is the phase mismatch given by:

$$\begin{aligned}\Delta k_{11} &= \beta_p^{(1)} - \beta_s^{(0)} - \beta_i^{(1)} - K \\ \Delta k_{12} &= \beta_p^{(1)} - \beta_s^{(1)} - \beta_i^{(0)} - K \\ \Delta k_{13} &= \beta_p^{(1)} - \beta_s^{(1)} - \beta_i^{(2)} - K \\ \Delta k_{14} &= \beta_p^{(1)} - \beta_s^{(2)} - \beta_i^{(1)} - K \end{aligned} \quad (22)$$

$I_{1j}$ = overlap integrals between the pump, signal and idler of $j$=1, 2, and 3 processes, and its expression is given by:

$$\begin{aligned}I_{11} &= \int u_p^{(1)}(\vec{r})u_s^{(0)}(\vec{r})u_i^{(1)}(\vec{r})dydz \\ I_{12} &= \int u_p^{(1)}(\vec{r})u_s^{(1)}(\vec{r})u_i^{(0)}(\vec{r})dydz \\ I_{13} &= \int u_p^{(1)}(\vec{r})u_s^{(1)}(\vec{r})u_i^{(2)}(\vec{r})dydz \\ I_{14} &= \int u_p^{(1)}(\vec{r})u_s^{(2)}(\vec{r})u_i^{(1)}(\vec{r})dydz \end{aligned} \quad (23)$$

The efficiency of the four down conversion processes for $j$=1, 2, 3 and 4 are proportional to $|F_{1j}|^2$.

In Figure 2, we have shown that the three waveguides are again recombined to a single multimode waveguide supporting three normal modes. As the transformation from the triplet of channels to the output three moded waveguide is unitary, the fundamental symmetric (0), the first excited antisymmetric (1) and the second excited symmetric (2) normal modes at signal and idler wavelengths would respectively excite the fundamental first, excited antisymmetric and second excited symmetric modes of three–moded waveguide. Thus in the three moded single waveguide region, the modal entanglement would be preserved but now instead of in terms of the normal modes of three coupled waveguides, it will be modal entanglement among the three normal modes of the output three moded waveguide.

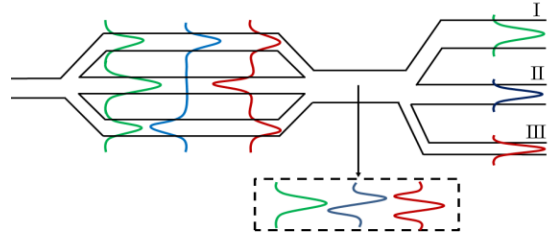

FIG. 3: (Color online) Schematic of waveguide structure describing the how the modes are separated at output.

If required, the three modes can be spatially separated into three different output waveguides by using an asymmetric three waveguide splitter in which all the three waveguide are single-moded but with different propagation constants (which can be easily obtained by choosing different widths of the three waveguides). In such a device which is an extension of the asymmetric Y-splitter that is used in integrated optics, the fundamental symmetric mode at both the signal and the idler wavelengths will exit from the waveguide having highest propagation constant (upper most waveguide in Figure 3), the second excited symmetric modes will exit from the waveguide having the lowest propagation constant (lowest waveguide in Figure 3) while the first excited antisymmetric mode will exit from the middle waveguide having propagation constant in between the other two waveguides. This way the modes in which the photons are generated can be separated spatially leading to path entangled photon pairs.

If we number the output waveguides shown in Figure 3 as I, II and III, then the output quantum state for the first case would be described by:

$$|\psi_0\rangle = i\int d\omega\,[C_{01}|H_{sI},V_{iI}\rangle + C_{02}|H_{sII},V_{iII}\rangle + C_{03}|H_{sIII},V_{iIII}\rangle] \quad (24)$$

And for the second case it would be described by:

$$|\psi_1\rangle = i\int d\omega\,[C_{11}|H_{sI},V_{iII}\rangle + C_{12}|H_{sII},V_{iI}\rangle + C_{13}|H_{sII},V_{iIII}\rangle + C_{14}|H_{sIII},V_{iII}\rangle] \quad (25)$$

Each mode of the intermediate three moded waveguide supporting three normal modes would have different field profiles as shown in Figure 3. When the photons are separated into three output waveguides, each of the output waveguide will have an intensity profile characteristic of that waveguide and the quantum state of the output state would be described by Eq. (24) and (25) from which the probability of finding the photons in any of the waveguides with a particular polarization can be easily determined. It may be also worth mentioning here that the photons are separated at the output by their polarizations. That is, polarization defines two subsystems (we have one photon in each polarization), and sets of mutually orthogonal spatial modes define Hilbert spaces of individual subsystems.

## IV. NUMERICAL SIMULATIONS

In order to show the feasibility of the idea, we consider generation of entangled photons in a directional coupler device using titanium indiffused lithium Niobate channel waveguide. For the analysis we assume the waveguides to have a step index profile. The value of the lithium niobate substrate refractive index ($n_s$) for different wavelengths and different polarizations were calculated using Sellmeier equation given in Ref [19]. The refractive index difference ($\Delta n$) for a waveguide has been calculated using Ref [20].

We have carried out numerical simulations and optimization of the waveguide parameters assuming a pump wavelength of 675 nm.

We consider the three identical channel waveguide of width = 6 µm and depth = 7 µm separated by a distance of 6 µm. The propagation constant of normal modes of three waveguide directional coupler can be obtained by solving the Eigen value equation which is obtained by writing the fields in all regions and applying appropriate boundary conditions. After calculating the propagation constant of normal modes at pump, signal and idler wavelengths, we can obtain the field distribution of these modes. Details are given in the Appendix.

### A. Pump is in the fundamental symmetric mode

Figure 4 shows the transverse electric field pattern of the three normal modes along *y*-direction at pump, signal and idler for the three processes involved when the H-polarized fundamental symmetric mode is incident on waveguide at pump wavelength of 675nm.

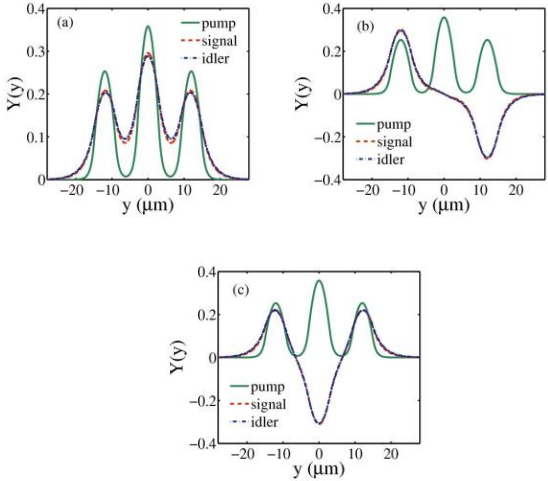

FIG. 4: (Color online) Normalized transverse modal field distributions along y - direction in a three waveguide coupler (a) fundamental symmetric normal mode (0) at pump, signal and idler wavelengths; (b) fundamental symmetric mode normal mode (0) at pump wavelength and the first excited anti symmetric normal mode (1) at signal and idler wavelengths; (c) fundamental symmetric mode normal mode (0) at pump wavelength and the second excited symmetric normal mode (2) at signal and idler wavelengths

Figure 5 shows the measurable square of the corresponding fields.

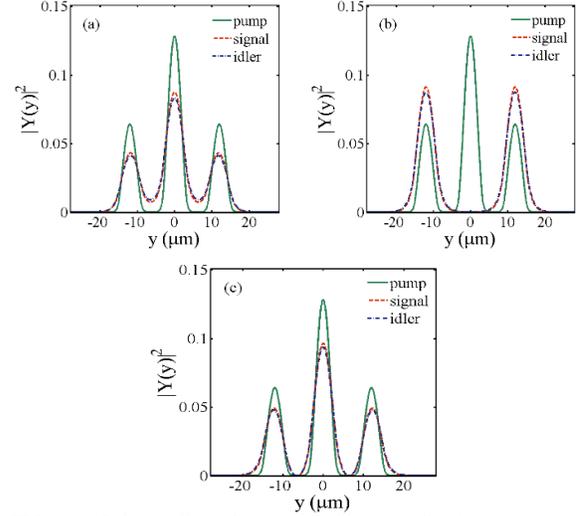

FIG. 5: (Color online) Square of the normalized transverse modal intensity distributions along y - direction in a three waveguide coupler (a) fundamental symmetric normal mode (0) at pump, signal and idler wavelengths; (b) fundamental symmetric mode normal mode (0) at pump wavelength and the first excited anti symmetric normal mode (1) at signal and idler wavelengths; (c) fundamental symmetric mode normal mode (0) at pump wavelength and the second excited symmetric normal mode (2) at signal and idler wavelengths

For waveguide width of $a = 6$ µm, depth = 7 µm and separation $d = 6$µm, the overlap integral of all three process are almost equal and the grating spatial frequency required is 0.9074 µm$^{-1}$. Figure 6 shows the variation of down conversion efficiency vs signal wavelength and QPM grating for $L = 2.55$ mm for all three process. From Figure 6(a), it can be seen that the three curves intersect at a signal wavelength of 1350 nm. Thus by using a narrow band wavelength filter at the signal wavelength 1350 nm, we can obtain higher dimensional mode entangled photon pairs at the output. The role of the filter is two-fold: Not only it is used to select the point where equal conversion efficiencies are achieved, but also to remove spectral correlations between the signal and idler photons, which could lead to a significant reduction of entanglement quality [21].

In order to show the dependence on the QPM grating period, in Figure 6(b) we have plotted the variation of efficiency with QPM grating spatial frequency *K* keeping the signal and idler wavelength to be 1350 nm.

It can be seen from Figure 6(b) that at QPM grating spatial frequency $K = 0.9074$µm$^{-1}$ all the three processes intersect and have same efficiency.

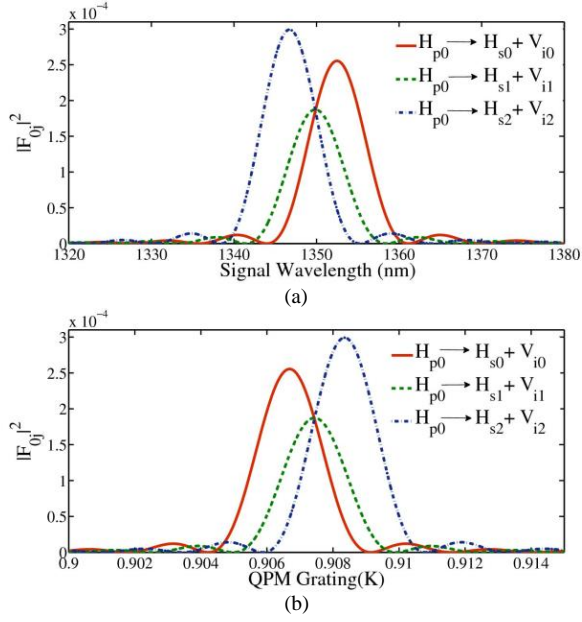

FIG. 6: (Color online) Variation of the efficiency of the three down conversion processes (a) as a function of signal wavelength; (b) as a function of QPM grating.

Thus, using a single grating of grating period $\Lambda = 6.92$ μm in three waveguide coupler and an appropriate wavelength filter, we can obtain high dimensional entangled photon pairs at output.

We have found that for an error in the grating period of ± 50 nm, the signal wavelength changes by ± 5 nm and to achieve degeneracy we may tune the pump wavelength by ±2.5nm to obtain the maximally entangled state. Figure 7 shows the variation of the efficiency of the three possible down conversion processes as a function of the signal wavelength with slightly different grating periods.

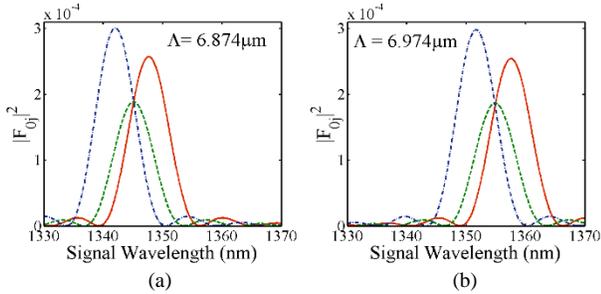

FIG. 7: (Color online) Variation of the efficiency of the three down conversion processes as a function of signal wavelength having grating period (a) $\Lambda$ =6.874 μm; (b) $\Lambda$= 6.974μm

As can be seen from Figure 7, the curves still intersect at one value of signal wavelength which changes with the period. However choosing an appropriate wavelength filter can provide us with mode entangled degenerate pairs of photons even with slight errors in the grating period. Thus small errors in the grating period can be taken into account by a small tuning of the pump wavelength so that degenerate mode entangled photon pairs are generated.

## B. Pump is in the first excited antisymmetric normal mode

When H-polarized first excited anti symmetric normal mode is incident on a waveguide at pump wavelength of 675 nm then it down convert via the four processes as described in Sec. II through a degenerate type II SPDC process. The square of normal modes of three waveguide coupler involved in these processes are shown in Figure 8.

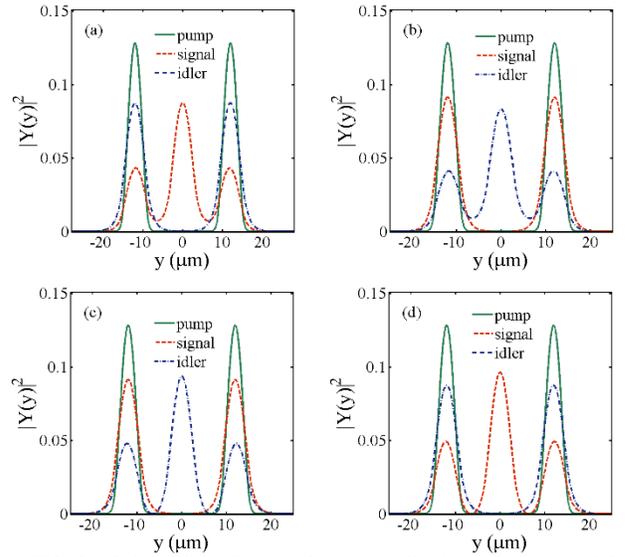

FIG. 8: (Color online) Square of the normalized transverse modal field distributions along y –direction in a three waveguide coupler (a) first excited antisymmetric normal mode (1) at pump and idler wavelength and the fundamental symmetric normal mode (0) at signal wavelengths in a three waveguide coupler; (b) first excited antisymmetric normal mode (1) at pump and signal wavelength and the fundamental symmetric normal mode (0) at idler wavelengths; (c) first excited antisymmetric normal mode (1) at pump and signal wavelengths and the second excited symmetric normal mode (2) at idler wavelength (d) first excited antisymmetric normal mode (1) at pump and idler wavelengths and the second excited symmetric normal mode (2) at signal wavelength

In this case also, with the same waveguide parameters and length of crystal, the overlap integral of all the four processes become almost equal. Figure 9 shows the variation of down conversion efficiency vs signal wavelength and QPM grating for $L = 2.55$ mm for all three process. In this case also, it can be seen from Figure 9 (a) that the four curves intersect at a signal wavelength of 1350 nm, but have the broader spectrum as compared to case 1. Thus by using a wavelength filter around the signal wavelength 1350 nm, we can obtain partial higher dimensional mode entangled photon pairs at the output.

In order to show the dependence on the QPM grating period, in Figure 9(b) we have plotted the variation of efficiency with QPM grating spatial frequency $K$ keeping the signal and idler wavelength to be 1350 nm.

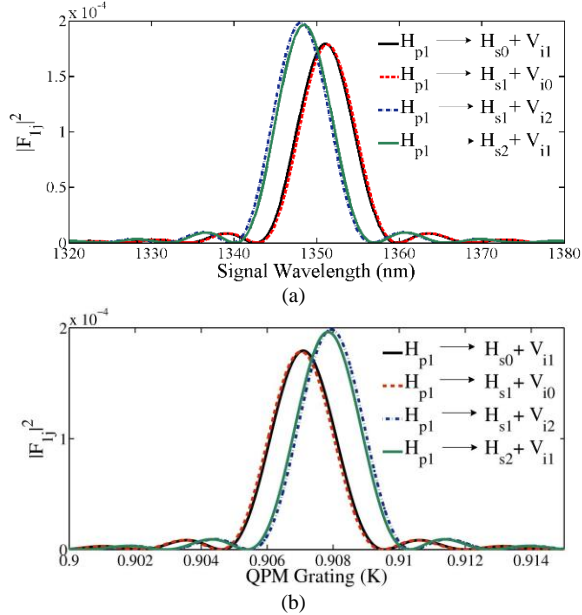

Fig. 9: (Color online) Variation of the efficiency of the four down conversion processes (a) as a function of signal wavelength; (b) as a function of QPM grating

It can be seen from Figure 9(b) that at QPM grating spatial frequency $K = 0.9074 \mu m^{-1}$ all the four processes intersect and have same efficiency. Thus, using a single grating of grating period $\Lambda = 6.92 \mu m$ in three waveguide coupler, we can obtain partial high dimensional entangled photon pairs at output.

We would like to mention that the modal entanglement properties of the photon pair could be demonstrated by full quantum state tomography applied to the spatial degree of freedom. One could use for example measurement of the Wigner function demonstrated in Ref [22]. This scheme has direct extension to more than one photon and could be used to reconstruct the complete quantum state of the two-photon system.

Another route could be to use a carefully selected set of mutually non-orthogonal bases which would demonstrate entanglement through e.g. violation of Bell's inequalities. This has been done in the case of orbital angular momentum [11, 23] and could be extended to other sets of modes with the help of spatial light modulators.

## V. CONCLUSION

We have shown that by an appropriate design of a three waveguide directional coupler, it is possible to generate mode entangled state in a higher dimensional space by proper designing of the device. Also with the same device and having the same grating period we can obtain the partial high dimensional entanglement with more tolerance. The advantage of using this waveguide structure is the flexibility and the design space availability to achieve desired characteristics of the photon pairs generated in the down conversion process. Further investigation on coupled waveguide structures are expected to provide the possibility of designing waveguide devices generating a larger range of entangled and hyper entangled photon pairs.

## ACKNOWLEDGEMENTS

The authors wish to thank Dr. Rafal Demkowicz-Dobrzanski, Faculty of Physics, University of Warsaw, Poland for technical discussions. This research was partially supported by the "Polish NCBiR under the ERA-NET CHIST-ERA project QUASAR".

## APPENDIX-A

Consider a three waveguide coupler as shown in Figure 10(a) and its corresponding separable refractive index profile is shown in Figure 10(b).

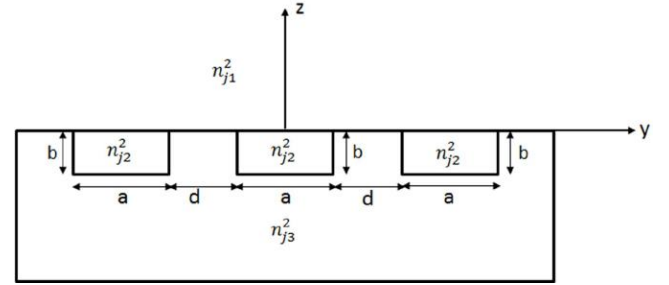

FIG. 10(a): Schematic of three waveguide coupler

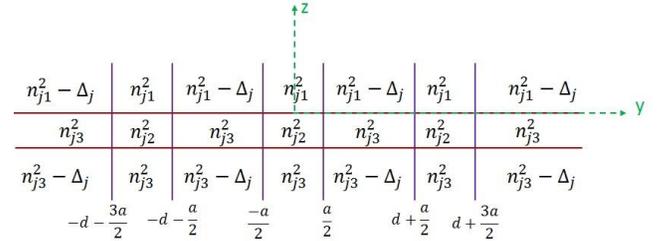

FIG.10 (b): Corresponding separable refractive index profile

The refractive index profile corresponding to above structure can be written as [24]:

$$n_j^2(y,z) = n_j^2(y) + n_j^2(z) - n_{j2}^2 \quad ; j = y, z$$

where,

$$n_j^2(y) = \begin{cases} n_{j2}^2 & ; \left(d+\frac{a}{2}\right) < |y| < \left(d+\frac{3a}{2}\right) \ \& \ |y| < \left(\frac{a}{2}\right) \\ n_{j3}^2 & ; \left(\frac{a}{2}\right) < |y| < \left(d+\frac{a}{2}\right) \ \& \ |y| > \left(d+\frac{3a}{2}\right) \end{cases}$$

$$n_j^2(z) = \begin{cases} n_{j1}^2 & ; z > 0 \\ n_{j2}^2 & ; -b < z < 0 \\ n_{j3}^2 & ; z < -b \end{cases}$$

The above profile is correctly represent the refractive index profile of three channel waveguide coupler in every region except in the corner regions where the two profiles differ by an amount of $\Delta_j = n_{j2}^2 - n_{j3}^2$, which is very small, so for correction in corner region we have used the Perturbation method [24].

The transverse modal field profile in the three waveguide coupler is given by:

$$u_j^{(mn)}(r) = Y_j^{(m)}(y) Z_j^{(n)}(z)$$

where,

$Y_j^{(m)}(y)$ is the field pattern of the $j$– polarized $m^{th}$ mode corresponding to the waveguide structure along y-direction and is given by:

$$Y_j^{(m)}(y) = \begin{cases} A_j^m e^{-\gamma_j^m |y|} & ; |y| \geq \left(d+\frac{3a}{2}\right) \\ B_j^m \cos(\kappa_j^m |y|) \\ + C_j^m \sin(\kappa_j^m |y|) & ; \left(d+\frac{a}{2}\right) \leq |y| \leq \left(d+\frac{3a}{2}\right) \\ D_j^m \cosh(\gamma_j^m |y|) \\ + E_j^m \sinh(\gamma_j^m |y|) & ; \left(\frac{a}{2}\right) \leq |y| \leq \left(d+\frac{a}{2}\right) \\ F_j^m \begin{pmatrix} \cos(\kappa_j^m y) \\ \sin(\kappa_j^m y) \end{pmatrix} & ; |y| \leq \left(\frac{a}{2}\right) \end{cases}$$

and $Z_j^{(n)}(z)$ is the field pattern of the $j$– polarized $n^{th}$ mode corresponding to the waveguide structure along z-direction and is given by:

$$Z_j^{(n)}(z) = \begin{cases} G_j^n e^{(-\eta_j^n z)} & ; z \geq 0 \\ H_{j1}^n \cos(\sigma_j^n z) + H_{j2}^n \sin(\sigma_j^n z) & ; -b \leq z \leq 0 \\ J_j^n e^{(\delta_j^n z)} & ; z \leq -b \end{cases}$$

where,

$$\kappa_j^m = \sqrt{k_0^2 n_{j2}^2 - (\beta_{jy}^m)^2}, \quad \gamma_j^m = \sqrt{(\beta_{jy}^m)^2 - k_0^2 n_{j3}^2}$$

$$\delta_j^n = \sqrt{(\beta_{jz}^n)^2 - k_0^2 n_{j3}^2}, \quad \eta_j^n = \sqrt{(\beta_{jz}^n)^2 - k_0^2 n_{j1}^2}$$

$$\sigma_j^n = \sqrt{k_0^2 n_{j2}^2 - (\beta_{jz}^n)^2}$$

$$k_0 = \frac{2\pi}{\lambda}$$

A, B, C, D, E, F, G, H$_1$, H$_2$ and J are constants which can be find by applying appropriate boundary conditions at the interface, $\beta_{jy(z)}^{m(n)}$ is the propagation constant of the $j$–polarized $m^{th}(n^{th})$ mode corresponding to waveguide structure along the $y^{th}(z^{th})$ - direction

The propagation constant of the $j$– polarized $(mn)$ mode of the complete structure is given by [24]:

$$\left(\beta_j^{mn}\right)^2 = \left(\beta_{jy}^m\right)^2 + \left(\beta_{jz}^n\right)^2 - k_0^2 n_{j2}^2 + \Delta\beta$$

where, $\Delta\beta_j = k_0^2 \dfrac{\int\limits_{-\infty}^{\infty}\int\limits_{-\infty}^{\infty} \Delta_j \left|u_j^{(mn)}(y,z)\right|^2 dydz}{\int\limits_{-\infty}^{\infty}\int\limits_{-\infty}^{\infty} \left|u_j^{(mn)}(y,z)\right|^2 dydz}$

The y-polarized field would corresponds to a TM mode in the y- direction and TE mode in the z-direction whereas the z-polarized field would corresponds to a TE mode in the y-direction and TM mode in the z-direction and obtained the propagation constant by solving corresponding Eigen value equations.

Thus, for calculating the propagation constant of y-polarized $mn$ mode, we solve following set of Eigen value equation of TM mode along the y-direction and TE mode along the z-direction.

$$\frac{\kappa_y^m}{\gamma_y^m} \tan\left(\frac{\kappa_y^m a}{2}\right) = \left(\frac{n_{y2}}{n_{y3}}\right)^2 \left[\frac{\tanh(\gamma_y^m d) - \dfrac{t_{1y}^m}{\gamma_y^m}}{1 - \left(\dfrac{t_{1y}^m}{\gamma_y^m}\right)\tanh(\gamma_y^m d)}\right]$$

$$-\frac{\kappa_y^m}{\gamma_y^m} \cot\left(\frac{\kappa_y^m a}{2}\right) = \left(\frac{n_{y2}}{n_{y3}}\right)^2 \left[\frac{\tanh(\gamma_y^m d) - \dfrac{t_{1y}^m}{\gamma_y^m}}{1 - \left(\dfrac{t_{1y}^m}{\gamma_y^m}\right)\tanh(\gamma_y^m d)}\right]$$

$$\tan(\delta_y^n b) = \frac{\left(\frac{\delta_y^n}{\sigma_y^n}\right) + \left(\frac{\eta_y^n}{\sigma_y^n}\right)}{1 - \left(\eta_y^n/\sigma_y^n\right)\left(\delta_y^n/\sigma_y^n\right)}$$

where, $t_{1y}^m = \kappa_y^m \left(\frac{n_{y3}}{n_{y2}}\right)^2 \left[\frac{\left(\frac{\kappa_y^m}{\gamma_y^m}\right)\tan(\kappa_y^m a) - \left(\frac{n_{y2}}{n_{y3}}\right)^2}{\left(\frac{\kappa_y^m}{\gamma_y^m}\right) + \left(\frac{n_{y2}}{n_{y3}}\right)^2 \tan(\kappa_y^m a)}\right]$

$$\tan(\delta_z^n b) = \frac{\left(\frac{n_{z2}}{n_{z3}}\right)^2 \left(\frac{\delta_z^n}{\sigma_z^n}\right) + \left(\frac{n_{z2}}{n_{z1}}\right)^2 \left(\frac{\eta_z^n}{\sigma_z^n}\right)}{1 - \left(\frac{n_{z2}}{n_{z3}}\right)^2 \left(\frac{\eta_z^n}{\sigma_z^n}\right)\left(\frac{n_{z2}}{n_{z1}}\right)^2 \left(\frac{\delta_z^n}{\sigma_z^n}\right)}$$

where, $r_{1z} = \kappa_z^m \left[\frac{\kappa_z^m \tan(\kappa_z^m a) - \gamma_z^m}{\kappa_z^m + \gamma_z^m \tan(\kappa_z^m a)}\right]$

Now, for calculating the propagation constant of z-polarized field, we solve following set of Eigen value equation corresponding to TE mode along y-direction and TM mode along z-direction.

$$\frac{\kappa_z^m}{\gamma_z^m} \tan\left(\frac{\kappa_z^m a}{2}\right) = \left[\frac{\tanh(\gamma_z^m d) - \frac{r_{1z}^m}{\gamma_z^m}}{1 - \left(\frac{r_{1z}^m}{\gamma_z^m}\right)\tanh(\gamma_z^m d)}\right]$$

$$-\frac{\kappa_z^m}{\gamma_z^m} \cot\left(\frac{\kappa_z^m a}{2}\right) = \left[\frac{\tanh(\gamma_z^m d) - \frac{r_{1z}^m}{\gamma_z^m}}{1 - \left(\frac{r_{1z}^m}{\gamma_z^m}\right)\tanh(\gamma_z^m d)}\right]$$